# An improved high-resolution shear velocity model beneath the Iranian plateau using adjoint noise tomography

Abolfazl Komeazi[1*], Ayoub Kaviani[1], Farzam Yaminifard[2], Mohammad Tatar[2], Georg Rümpker[1,3]

1. Institute of Geosciences, Goethe-University Frankfurt, Frankfurt, Germany

2. International Institute of Earthquake Engineering and Seismology (IIEES), Tehran, Iran

3. Frankfurt Institute for Advanced Studies, Frankfurt, Germany

**Abstract**

We perform an adjoint waveform tomography using Rayleigh wave Empirical Green's Functions (EGFs) at 10-50 s periods to improve a prior 3-D velocity model of the crust and uppermost mantle beneath the Iranian Plateau. EGFs were derived from cross-correlations of ~8 years of continuous vertical component seismic noise recorded by 119 broadband stations within the region. Adjoint tomography refines the initial model by iteratively minimizing the frequency-dependent travel-time misfits between the synthetic Green's Functions (SGFs) and EGFs measured in different period bands. The total misfit is dropped by ~75 percent after 6 iterations. Overall, the adjoint tomography provides images with more realistic and improved resolutions and amplitudes due to the inclusion of finite-frequency waveforms. The use of a numerical spectral-element solver in adjoint tomography provides highly precise structural sensitivity kernels, resulting in more robust images compared to those generated by ray-theory tomography. Our study also demonstrates improvement in lateral resolution and depth sensitivity. The final model adjusts the shapes of velocity anomalies at crustal and uppermost mantle depths especially in the Zagros convergence zone. These improvements include a high-resolution image of delamination of the subducting Arabian lithosphere from Zagros lower crust beneath the central and NW Zagros. The refine, high-resolution images also better show the geometry of low-velocity anomalies, which represent diachronous underthrusting of Arabian crust beneath central Iran.

Keywords: spectral element method, adjoint tomography, Iranian plateau, ambient noise, Zagros collision zone.

## Introduction

The long-lasting collision between the Arabian plate and Eurasia, after the closure of the Neo-Tethys around ~35 Ma, resulted in underthrusting of the rifted Arabian lithosphere beneath the Eurasian plate and consequently significant lithospheric deformation in the Iranian plateau (e.g., Mouthereau et al., 2012, and references therein). Induced stress resulting from the convergence is mostly accommodated in the form of thickening/shortening beneath the Zagros Fold and Thrust Belt (ZFTB) and Alborz and Kopet-Dagh mountain ranges (e.g., Priestley et al., 2012), as well as by lithospheric-scale displacements along the transcurrent faults within the less deformed microplates in Central Iran (CI). This is also confirmed by recorded seismic activity (Fig. 1) and the GPS measurements in the study area (e.g., Vernant et al., 2004). Lithospheric deformation in the region is investigated by various geological and geophysical studies (e.g., Masson et al., 2005; Omrani et al., 2008; Hatzfeld et al., 2010; Agard et al., 2005, 2011), including tomography investigations (e.g., Kaviani et al., 2007; Paul et al., 2010; Shad Manaman et al., 2011; Al-lazki et al., 2014; Motaghi et al., 2017a,b; Kaviani et al., 2020; Irandoust et al., 2022). In general, using different techniques, these studies emphasize the presence of high-velocity and thick lithosphere beneath the ZFTB, Sanandaj-Sirjan metamorphic Zone (SSZ) and Urumieh–Dokhtar Magmatic Arc (UDMA), and low-velocity and thin lithosphere beneath central Iran. With advancements in data collection and computational resources, tomographic images across the Iranian plateau have significantly improved over the last decade. Nevertheless, these images are subject to simplifying assumptions (such as high-frequency frequency assumptions) with limited resolution, which leave the geometry of the crust and the underthrusting slab in the Zagros collision zone a topic of ongoing debate.

The successful implementation of ambient noise tomography by Shapiro et al. (2005) was followed by many other studies (e.g., Lin et al., 2008; Chen et al., 2014; Gao and Shen, 2014). The method is based on the extraction of surface wave Green's function between different receivers by cross correlation of long time ambient seismic noise to image Earth structures. This method has proven to be particularly effective for investigating crustal structures, offering enhanced resolution. Its utility becomes evident in Iran, where seismic station distribution is sparse, limiting the effectiveness of traditional earthquake travel-time tomography. Over the past decade, this method has been employed to investigate the crustal and uppermost mantle structure beneath Iran (e.g., Mottaghi et al., 2013; Ansaripour et al., 2019; Movaghari and Javan-Doloie, 2020; Pilia et al., 2020; Kaviani et al., 2020). However, using this technique with the ray-theory assumption may not accurately represent the complexities of subsurface structures, such as sharp velocity discontinuities. This can result in limited resolution and accuracy, particularly in our study area which is regarded as a complex geological region (e.g., Mouthereau et al., 2012; Agard et al., 2005). In recent decades, advancements in computational capabilities have opened up new possibilities for seismologists to employ accurate numerical methods in 3D full waveform tomography (Tromp et al., 2005; Liu and Tromp, 2006; Fichtner et al., 2009; Tape et al., 2010;

Liu and Gu, 2012). This has led to the exploration of more sophisticated approaches, such as Adjoint noise tomography (ANT), which leverages precise 3D numerical simulations like the spectral element method (SEM, Komatitsch and Vilotte, 1998; Komatitsch and Tromp, 1999). These methods have the potential to reveal velocity perturbations of up to 10–30%, surpassing the capabilities of classical tomographic methods (e.g., Fichtner et al., 2009; Tape et al., 2009; Liu and Gu, 2012; Zhu et al., 2012). ANT involves sensitivity kernel computations based on the adjoint state method, and it was first introduced by Tromp et al. (2005), taking advantage of supercomputing facilities. This method shows promise in improving the resolution of crustal and uppermost mantle images (e.g., Wang et al., 2018, 2021; Lu et al., 2020).

In this study, we apply ANT to improve the pre-existing model (Kaviani et al., 2020) derived from traditional ambient noise tomography, aiming at creating a high-resolution tomographic image of the crust and sub-crustal lithospheric structure. This, in turn, may provide valuable insights into the complex interactions between the converging tectonic plates in the study area.

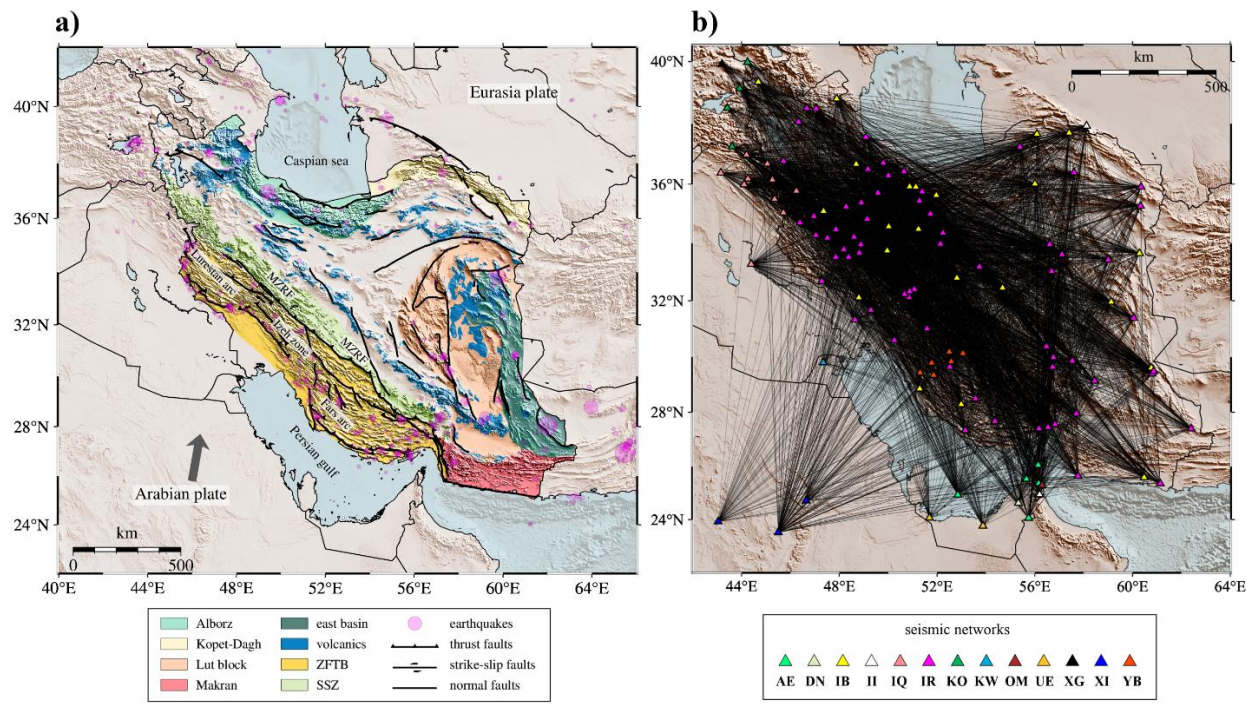

Figure 1. (a) Tectonic map featuring topography, bathymetry, and seismic activity within the study area. Seismic event data (earthquakes with a momentum magnitude greater than 5, since 1990) sourced from the NEIC catalog (https://earthquake.usgs.gov/earthquakes/search/). The colors are associated with various geological structures as indicated in the legend. (b) Location map of 119 seismic stations (triangles) and raypaths across the study area. Stations belonging to each specific network are represented using distinct colors.

## 2. Data and Method

The cross-correlations (CCs) employed in this investigation, serving as Empirical Green's Functions (EGFs), are obtained from a previous study conducted by Kaviani et al. (2020). A rigorous selection process resulted in the choice of approximately 3000 CCs for our study area, each having a signal-to-noise ratio (SNR) greater than 5. These CCs were derived from eight years of continuous seismic noise data recorded between 2007 and 2015, utilizing records from 119 broadband stations situated within the study area (Fig.1). The time series resulting from the CCs contain symmetric positive and negative legs. Therefore, to create a one-sided seismogram that can be directly compared to those generated by simulations involving a single station as the source and another as the receiver, we flip one side of these time series and then stack them. Since our analysis exclusively considers vertical components, the resulting set of cross-correlations is primarily characterized by Rayleigh wave signals. In the case of homogeneous noise source assumption, the obtained cross-correlations between two stations A and B are related to the EGFs as (Lobkis and Weaver, 2001):

$$-\frac{\partial C_{A,B}(t)}{\partial t} = G_{A,B}(t) \quad (1)$$

Where C is the noise cross correlation between points A and B, and G is the corresponding Green's function.

The initial velocity model used in this study is provided by Kaviani et al. (2020), which was established through conventional ambient noise tomography.

### 2.1 Setup and Implementation

We provide an overview of our implementation, focusing on the aspects associated with noise cross-correlation data. The underlying theory and concepts of adjoint noise tomography have been extensively covered in numerous prior studies (e.g., Luo and Schuster, 1991; Tromp et al., 2005; Virieux and Operto, 2009; Liu and Gu, 2012). Our workflow, depicted in Fig. 2, begins with the selection of high-quality CCs. From these CCs, we construct EGFs using Equation 1. By defining a mesh based on the initial velocity model (as illustrated in Fig.3), we are able to conduct forward simulations, computing SGFs at the receiver locations. Afterward, a proper window to capture the Rayleigh wave is selected and the misfit between the EGFs and SGFs is calculated. This misfit serves as the adjoint source, leading to the computation of misfit kernels. Further post-processing of these misfit kernels involves summation, smoothing, and preconditioning. At each iteration, we perform a line search to determine the optimal step length for updating the model. This process, starting from the forward simulation, is reiterated until the misfit values converge.

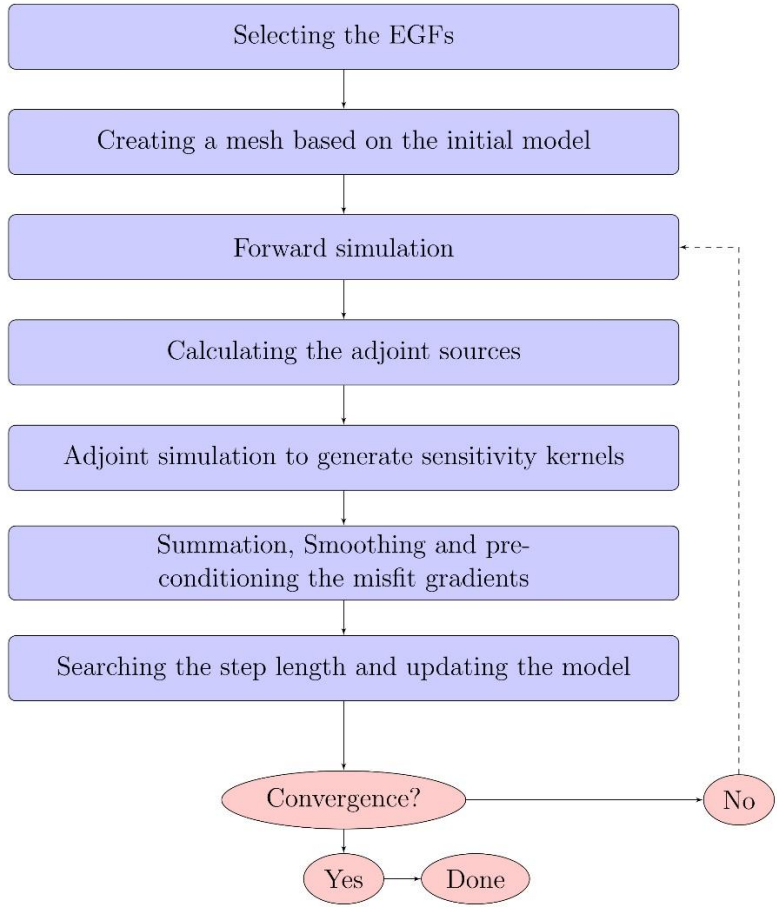

Figure 2. Workflow of the adjoint noise tomography.

The SEM has proven to be an accurate and efficient method of wavefield simulation in a 3-D heterogeneous media (Komatitsch and Vilotte, 1998; Komatitsch and Tromp, 1999, 2002a,b). It leverages the geometrical flexibility of finite-element methods and the exponential convergence rate of spectral methods. In this study, we utilize the open-source software package SPECFEM3D Cartesian (version 3) for both the forward and adjoint simulations. We also employ Cubit (version 15) to discretize our initial model using a mesh built in Cartesian coordinate. This mesh has 152 Elements longitude and 160 elements latitude and extends to a depth of 150 km with 10 elements (Fig. 3). The Gauss–Lobatto–Legendre (GLL) points are spaced at a minimum interval of approximately 2500 meters. These points serve as the nodes over which the partial derivative equations are solved. The configuration of this mesh gives sufficient simulation accuracy at periods equal and longer than 9.2 s. To optimize computational efficiency, we apply a bandpass filter ranging from 10 to 50 seconds to the signals, in contrast to the wider period band of 2 to 150 seconds applied to obtain the initial model. This choice reduces the computational burden by diminishing mesh dimensions and the number of required inversion iterations. Considering our target periods, we also disregard the attenuation effect due to its minor impact on our results.

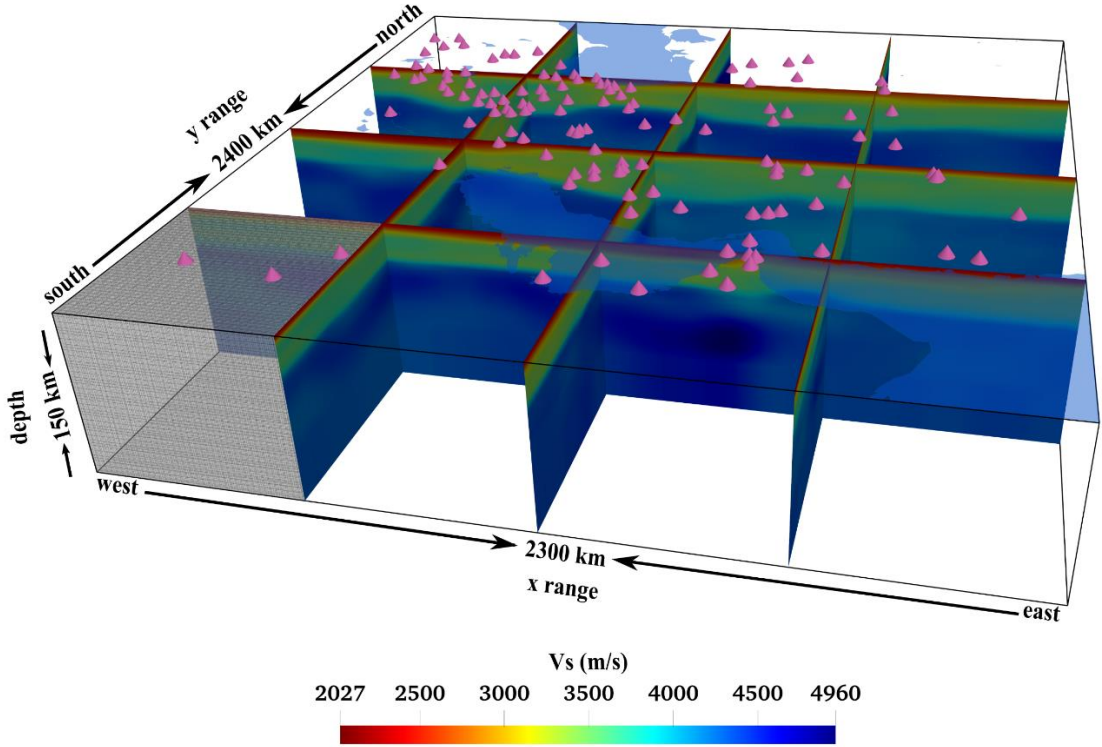

Figure 3. The mesh structure and shear wave velocity distribution of the initial model are shown. Shear wave velocity values of the initial model are displayed in six representative depth-sections, which also provide a schematic representation of the model's distribution across 12 GPU cores. A segment of the mesh for the southwest portion of the initial model is also illustrated. Seismic stations employed in this study are denoted by violet triangles, while the blue shading on top distinguishes between water and land.

To generate the SGFs in forward simulations, we put a single vertical point force $\hat{F} = f\hat{z}$ at the location of a station, which serves as a source, while all other stations act as receivers. We use a Gaussian function as the source time function for the point force. By this setup, the vertical component of the Green's Functions between the pairs of stations are calculated. The SEM computations are performed in parallel implementation using the Goethe-HLR GPU cluster (equipped with AMD Instinct MI210) using 12 cores. Each forward simulation typically takes about 5 min to generate seismograms of 500s duration, with a time step of 0.02.

Both the EGFs and SGFs are tapered to include only Rayleigh waves. For this purpose, we apply a time window defined as $[\frac{D}{V_{g(\max)}} - 25, \frac{D}{V_{g(\min)}} + 25]$, where D is the inter-station distance and $V_{g(\max)}$ and $V_{g(\min)}$ are maximum and minimum group velocities in the period range of 10-50 seconds, derived from Kaviani et al. (2020). In this study, we use the frequency dependent multi-taper travel-time difference as the misfit function (Tape et al., 2009):

$$\chi(m) = \frac{1}{2}\int W(\omega)[\tau^{obs}(\omega) - \tau(\omega,m)]^2 dm \quad (2)$$

Where $W$ denotes a windowing filter, the width of which depends on the frequency range within which the measurements are considered reliable. This width is also scaled by error estimates specific to that window. $\tau^{obs}(\omega)$ and $\tau(\omega,m)$ denote the observed and synthetic frequency-dependent travel-time for a single source-receiver pair. It is worth noting that during the stacking of noise cross-correlation, amplitude information in the data is lost. To address this, we normalize both the SGFs and EGFs to eliminate any potential amplitude effects in our calculations. We employ a series of control factors, such as minimum cross-correlation coefficient between the SGF and EGF, at each iteration to include only high-quality signals in the inversion process, as detailed in table S1. The final reported misfit value for each iteration is calculated as the average misfit value across all source-receiver pairs.

Following the computation of sensitivity kernels for all the sources (for a general review of sensitivity kernel calculations see Liu and Gu, 2012), along with the Hessian kernel, representing the second derivative of the misfit function, we sum these kernels. Subsequently, convolving these summed kernels with a gaussian function, we are able to remove artificial high wavenumber gradient artefacts (Trinh et al., 2017). In SPECFEM3D, these Hessian kernels approximate the diagonal elements of the Hessian matrix, since the accurate value of the Hessian matrix is not available or difficult to achieve. The approximation is determined as the vector inner product of the forward and adjoint accelerations as outlined below:

$$\tilde{H}(x) = \int \partial_t^2 s(x,t) \cdot \partial_t^2 s^\dagger(x, T-t) dt \quad (3)$$

where s and $s^\dagger$ are the forward and adjoint wavefields, respectively (Zhu et al., 2015). We use this approximate Hessian in the equation below to update the velocity model through a steepest descent gradient method:

$$m_{new} = m_{init} - \alpha \, \Delta m \quad (4)$$

where $\alpha$ is the step length we choose toward the minimum of the gradient for updating the model and $\Delta m = \nabla \chi \, \tilde{H}^{-1}$, representing the gradient of the misfit function $\chi$ (eq. 2) with respect to the model parameters, which is scaled by the inverse of the Hessian matrix. To determine the appropriate value of $\alpha$, we employ a line search strategy (Fig. S1). We set the $\alpha$ to values ranging from 1 to 3 percent with regard to the waveform misfits. Since the forward calculations for each trial step length can be computationally expensive, we select a specific set of sources with approximately uniform source-receiver ray-paths across the region (as illustrated in Fig. S2), to compute the misfits for each updated model with the respective step lengths. Finally, the step length corresponding to the minimum misfit value is selected to update the model at each iteration.

## 3. Results and Model validation

### 3.1 Horizontal cross-sections

In Fig. 4, we provide a comparative view of shear wave velocity depth sections at 15, 20, 30, 40, 50, and 60 km between the initial model (M00) and the final model (M12). A difference map between the M00 and M12 models are illustrated in Fig. S3. In this figure, the final model is masked based on sensitivity kernel values (Figs. S4 and S5), illustrating only the valid areas of the resolved model. Generally, perturbations relative to the mean Vs become more pronounced in the lower crust (30-40 km) in the new model. The low-velocity anomalies observed in the Zagros convergence zone coincident with the thickened crust in this region. High-velocity anomalies are primarily associated with the rigid central Iranian microplates, where the crust is reported to be thinner and, therefore, warmer. At depths of 20-30 km, the final model demonstrates significant disparities compared to the initial model, unveiling velocity anomalies with more detail. At these depths, new low-velocity features emerge in the Lurestan Arc and southern Zagros and Alborz is found to be non-homogeneous. The final model reveals high-velocity anomalies on the western and eastern sides of the Alborz mountain range, while the central (towards the west) parts are dominated by low velocities. While the initial model did detect velocity contrasts in the central and northeastern parts of the study area, these contrasts are much better pronounced in the final model. For example, the high-velocity anomaly in central Iran at a depth of 40 km is 10 percent higher in M12 compared to M00. Furthermore, there is a notable shift in the average Vs values, which increases in all the depth slices. This change in the average Vs is related to the shift in the mean travel-time misfit due to longer periods for model M00; a shift that is corrected in model M12 (Fig. 6). At the depth of 40 km, the shape of the low-velocity anomalies in the Zagros convergence zone and the northeastern area becomes narrower in the new model, and the high-velocity anomalies are more prominent. At depths greater than 50 km, the final model exhibits similar patterns to the initial model, with a slight increase in the average Vs values. In addition, some high-velocity features with more contrasts are observed in Central Iran, the Lut Block, and NE Iran.

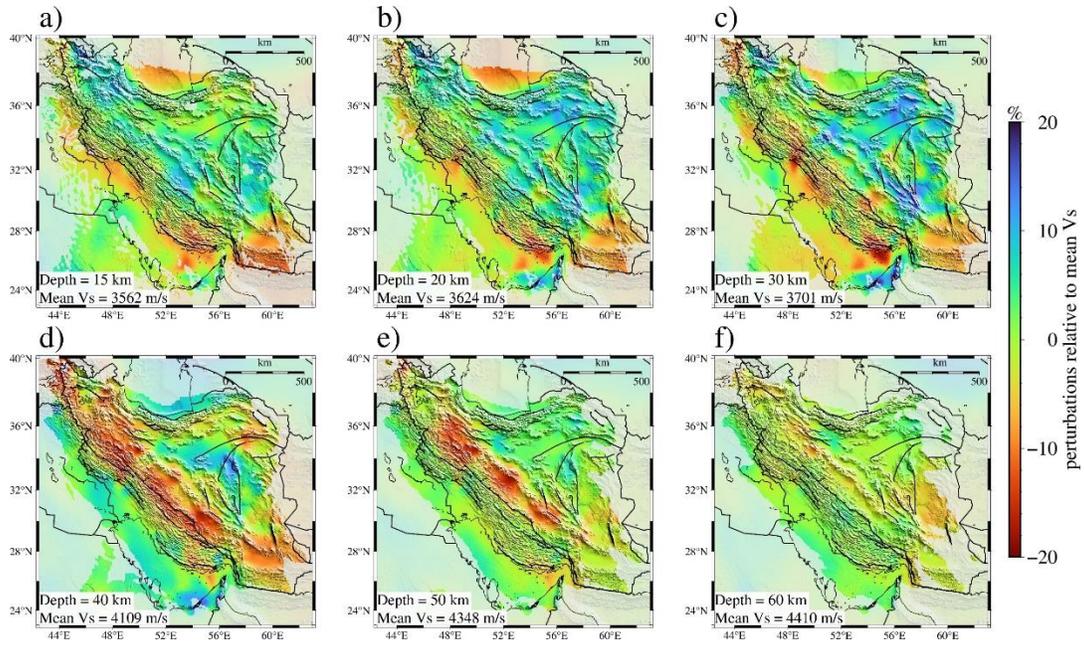

Figure 4. Comparison of shear wave velocity (Vs) depth slices at 15, 20, 30, 40, 50 and 60 km for both the final model (M12, panels a-f) and the initial model (M00, panels g-l). The relative velocity perturbation, expressed as a deviation from the mean velocity of each slice (values provided in the lower left corner of each map), is presented to highlight changes in the shape and amplitude of anomalies. The M12 depth slices (a-f) are masked based on the misfit kernel values (also refer to Fig. S4 and S5).

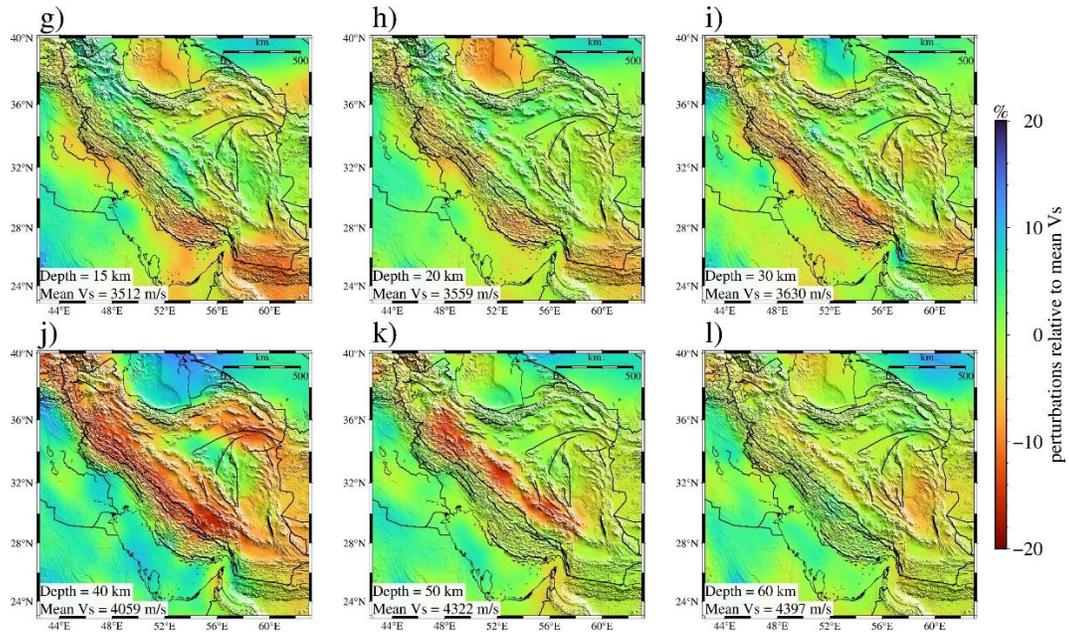

Figure 4. Continued.

## 3.2 Vertical cross-sections

In Fig. 5, we present four vertical sections from the final shear velocity model. The corresponding vertical sections from the initial model can be found in Fig. S6. Upon inspection of these figures, it becomes evident that the new model largely retains most of the pre-existing features from the initial model but with higher resolution and some refinements to the shape of the anomalies. We applied the same mask described in the previous section to exclude areas where model resolution is insufficient, ensuring model reliability. From south to north of Zagros (profiles A, B and C), a distinctive tongue-shaped low-velocity anomaly is mapped beneath the Zagros convergence zone. In profiles A, B, and C, some high-velocity anomalies are observed at shallow depths (approximately 20 km) in central Iran. With respect to the initial model, the tongue-shaped low-velocity anomaly has undergone modifications in the new model. For example, in profile A, this low-velocity anomaly does not extend continuously beneath the SSZ and UDMA, in contrast to the initial model. In profile C, a low-velocity corridor has emerged in the new model at subcrustal depths, beneath the SSZ and UDMA, which is absent in the initial model. In the same profile, a relatively high-velocity anomaly is accompanied by a local low-velocity anomaly beneath the southern Alborz region. The elongated low-velocity anomaly from the South Caspian Basin (SCB) to the southern Alborz in the initial model is truncated far north beneath the Alborz region. We have extracted the iso-velocity depth map corresponding to Vs = 4.2 km/s, as illustrated in Fig. S7. Assuming ~4.2 km/s as a proxy for crustal thickness, in comparison to the initial velocity model, the Moho depths across the region remain relatively consistent, with some adjustments around the Zagros collision zone and central Iran.

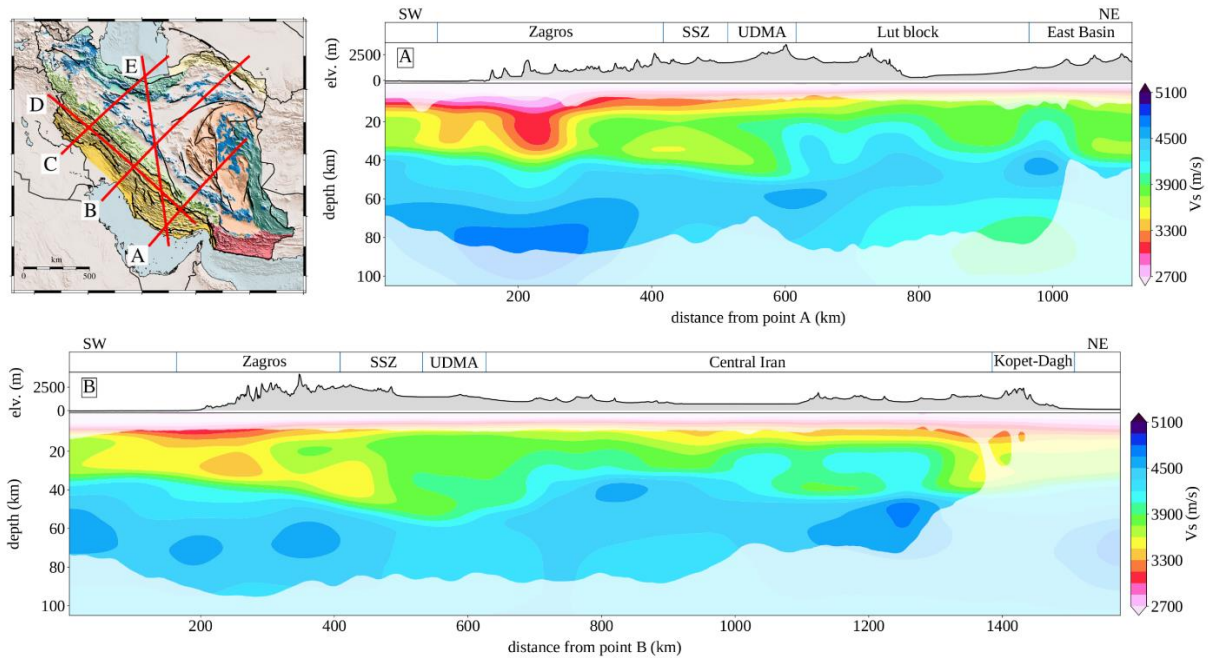

Figure 5. Vertical slices of the final shear wave velocity model (M12) along profiles A-E across the Iranian Plateau. The positions of these profiles are indicated on the map in the top-left corner. Vertical exaggeration: 4. Shear wave velocity values are masked according to the associated misfit kernel values (also refer to Figs. S4 and S5).

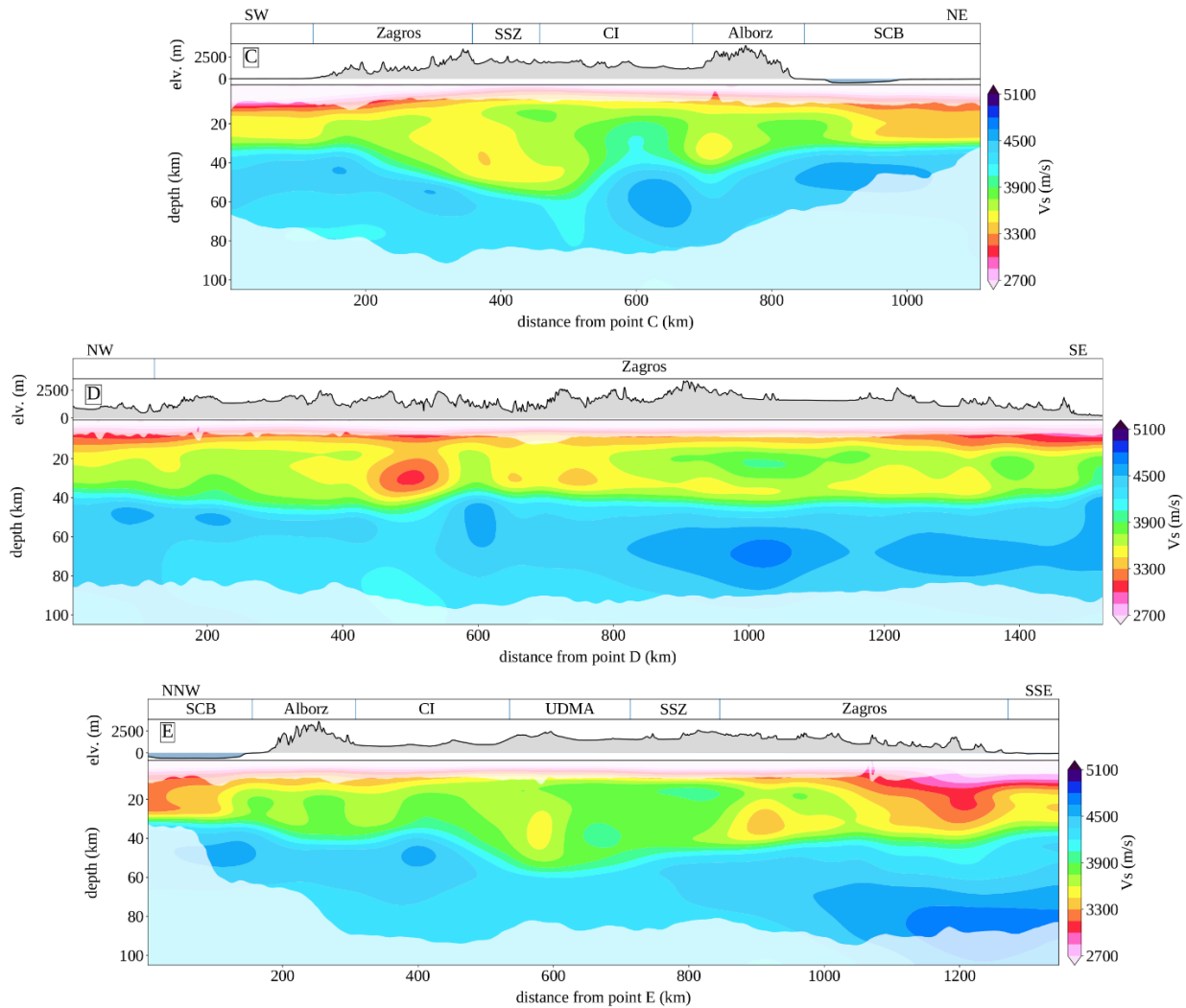

Figure 5. Continued.

## 3.3 Misfit reduction

The final model demonstrates a remarkable misfit reduction, approximately 75% relative to the initial model (Fig. 6a). In Fig. 6b, we provide a comparison of the travel time misfit histograms for both the initial and final models. Notably, the mean value of the misfits exhibits a distinct positive shift, indicating that the velocity values in the initial model tend to be lower than those in the true model. As previously discussed in Lu et al. (2020), traditional ambient noise tomography often exhibits this bias. Though there might not be a clear explanation for it, one possible factor in our study could be bias of the initial model towards higher periods, given that it was obtained within the period range of 2-150 seconds. Nevertheless, the final model demonstrates significant improvements in the histograms. The misfits in the final model are centered around zero and exhibit less standard deviation. Fig. 7 provides examples of observed waveforms and synthetics computed from both the initial and final models, highlighting a noticeable enhancement in waveform fitting at period bands 10-50, 10-20 and 20-50 seconds.

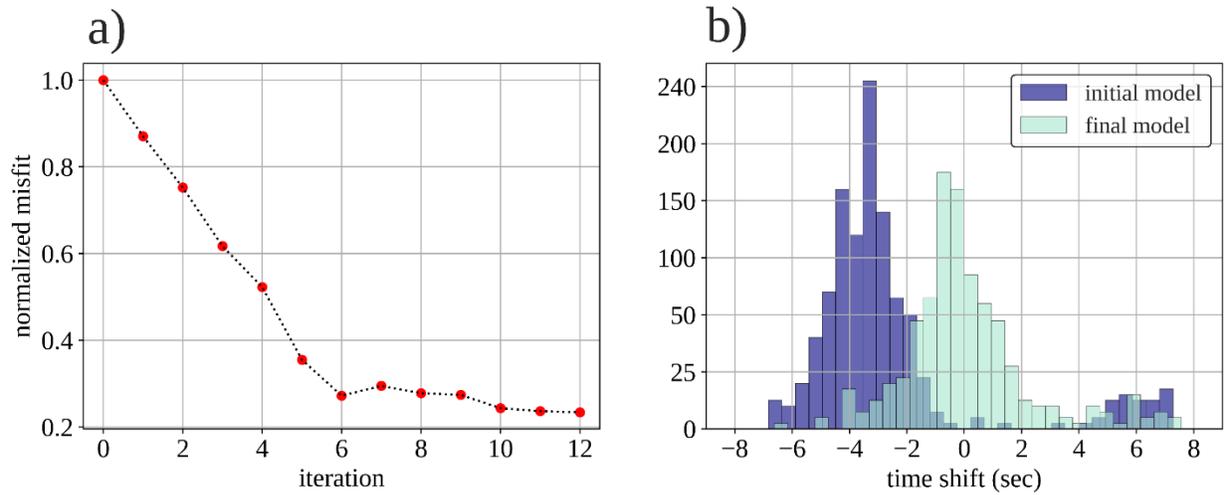

Figure 6: a) Total misfit reduction over iterations. b) Comparison of travel-time misfits between the final model (M12) and initial model (M00).

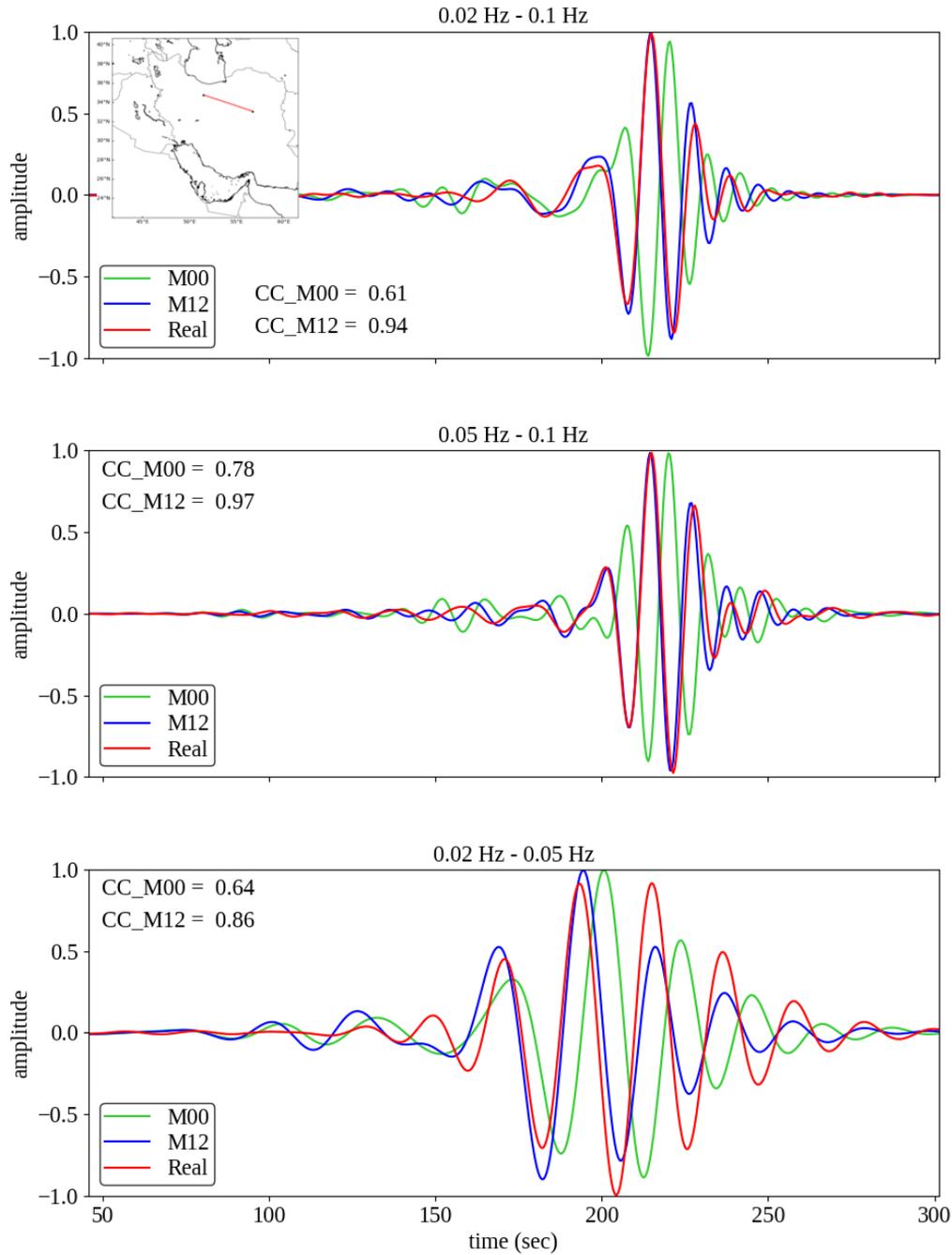

Figure 7. An example of waveform fitting for a source-receiver pair with 570 km distance is shown. In this graph the real waveform (EGF) with the waveforms resulting from the final model (M12, blue) and the initial model (M00, green) are compared. This comparison is done from top to bottom for period bands 10-50, 10-20, and 20-50 seconds, respectively. The value of the cross-correlation coefficient between the real and synthetic waveforms for both the initial and final model is denoted in the top left corner.

## 3.4 Resolution tests

Conducting traditional checkerboard tests to evaluate the quality of our 3D imaging is a resource-intensive process, demanding even more computational power compared to the actual structural inversion. This is due to the higher number of iterations required by the inversion process to converge. In two dimensions, such tests have been successfully executed, as demonstrated in previous study (Kaviani et al., 2020), concluding that the success of anomaly retrieval is primarily influenced by ray coverage. Therefore, in such cases, ray density maps serve as excellent indicators of resolution (Luo et al., 2013). In this study, to assess the resolution of our final model, we established a synthetic resolution test comparable to conventional checkerboard tests. As illustrated in Fig. 8a, we distributed 11 positive and negative velocity anomalies across the study area. These anomalies have the form of vertical cylinders with a radius of ~50 km, a depth thickness of ~20 km, and 10 percent of velocity perturbation relative to the initial model. To maintain similar resolution characteristics to the primary inversion process, we used the initial model (M00) as the background velocity model and introduced 10% of positive and negative anomaly to this background velocity model. We employed the same methodology described in Section 2 to generate synthetic waveforms, and subsequently, inverted them for the same source-receiver pair raypath coverage as in the observed data, following the workflow presented in Fig. 2. The results of the resolution test are presented in Fig. 8b and d, confirming that the velocity anomalies are relatively well-resolved along the Zagros collision zone. However, in the eastern and northeastern parts of the model, some smearing in the direction of Rayleigh wave propagation is observed.

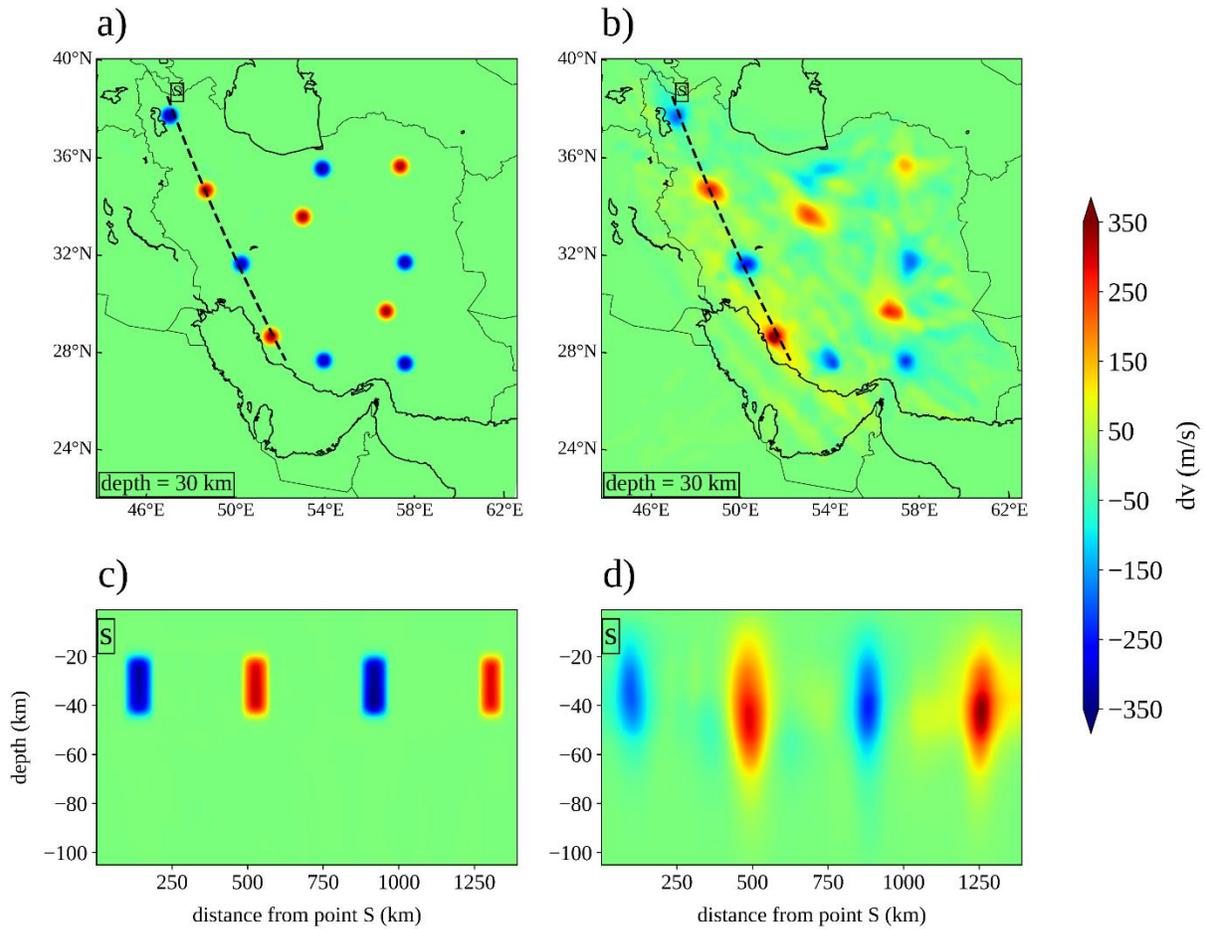

Figure 8. Evaluation of model resolution using a checkerboard-like test. Shown are 30 km depth slices and cross-sections of the initial model (a and c) and the recovered model (b and d) after subtracting the background model (M12). The black dashed line in (a) and (b) marks the location of the cross-section.

## 4. Discussion

By employing the highly accurate spectral element method to simulate the forward 3D wave propagation, coupled with the adjoint state method for addressing the inversion problem, we have successfully developed an improved, high-resolution 3D shear-wave velocity model for the crust and uppermost mantle across the Iranian Plateau. Our high-resolution maps, revealing more details, represent a significant enhancement relative to previous investigations (e.g., Kaviani et al., 2020; Motaghi et al., 2015; Shad Manaman et al., 2011; Shad Manaman and Shomali, 2010). The sensitivity kernel values (Fig. 5) indicate that based on the period range used in this study (10-50 sec), we are able to investigate the crustal and subcrustal structures, down to a depth of ∼70 km, across the Iranian Plateau. The horizontal maps (Fig. 4) demonstrate that the final model accurately represents the prominent geological structures in the region, encompassing the Zagros convergence zone, central Iran, and the Alborz-Kopet-Dagh system.

Shear wave velocities in the uppermost mantle consistently fall within the range of 4–4.3 km/s (see Figs. 4 and 5), comparable with other active tectonic zones within the Alp-Himalayan orogenic belt, such as the Anatolian Plateau (e.g., as discussed by Koulakov, 2011; Maggi and Priestley, 2005). Within the Zagros region, there is a persistent presence of elevated sub-Moho Vs values, indicating the existence of a high-velocity uppermost mantle layer beneath the Zagros convergence zone, as documented in numerous studies (Koulakov, 2011; Talebi et al., 2020; Mahmoodabadi et al., 2019, 2020; Priestley et al., 2012; Rahimi et al., 2014; Rahmani et al., 2019; Veisi et al., 2021, Irandoust et al., 2022). Moreover, the Vs variations along the Zagros (profiles A, B and C) provide valuable insights into the internal crustal structure within the collision zone.

In the Lurestan Arc (profile C), a distinctive, tongue-shaped low-Vs region (3.3–3.6 km/s) slopes toward the northeast. This feature initiates at shallow depths to the south of the Main Zagros Reverse Fault (MZRF) and extends to deeper levels in the crust beneath the Sanandaj-Sirjan Zone (SSZ) and the Urumieh–Dokhtar Magmatic Arc (UDMA), where the Moho reaches its maximum depth (~50 km). This thickened crustal root beneath the SSZ has been previously documented in geological (e.g., Agard et al., 2005), geophysical (e.g., Paul et al., 2006; Shad Manaman and Shomali, 2010; Motaghi et al., 2015, 2017a,b; Kaviani et al., 2007, 2020, Irandoust et al., 2022) and gravity (Snyder and Barazangi, 1986) studies. Comparing to the recent results of Irandoust et al. (2022), the slope of the tongue-shaped low-Vs in this region in our study is slightly steeper. It is noteworthy that this feature is observable in the initial model; however, in the new model, the shape and slope of the low-Vs anomalies are refined with higher resolution (Fig. S6). In central Zagros (profile B), this feature is also evident, albeit perhaps less prominently. In southern Zagros (profile A), the low velocity structure is different in comparison to the northern Zagros (profile C), extending to the northeast at shallower depths with a low angle, reaching well beneath the UDMA, where it exhibits a relatively sharp boundary with high velocities beneath the Lut block. In this profile, a notable low velocity zone beneath southern Zagros is detected. This could be attributed to a confined thickened sediment layer in the southeastern Zagros region, consistent with the narrower folding zone compared to the Fars arc, with GPS measurements indicating a similar convergence rate in both of these regions (Khorrami et al., 2019). The difference in crustal thickness between the northwest and southeast regions of Zagros was initially identified by Paul et al. (2010). However, in contrast to Paul et al. (2010), our high-resolution images better highlight this contrast by steeper inclination of the dipping low-velocity lower-crustal anomaly in the northwest compared to the southeast. The crustal thickening is more extensive in the southeast, as was also reported by Shadmanamen et al. (2011) and Irandoust et al. (2022).

The observed lateral variations in the structure of the leading edge of the advancing Arabian plate beneath central Iran (Fig. 5) provide compelling evidence for the proposed diachronous underthrusting process, as discussed in previous studies (e.g., Agard et al., 2005, 2011; Zhang et al., 2017). According to these investigations, the initiation of continental collision in the Zagros region commenced in the Oligocene in the northwest and progressed to southeastern Iran in the Miocene (e.g., Pirouz et al., 2017). GPS surveys conducted across the Zagros belt (e.g., Vernant et al., 2004; Walpersdorf et al., 2006; Khorrami et al., 2019) also reveal an increase in convergence rates along the Zagros region, with values escalating from the northwest (~4-6 mm/yr) to the southeast (~6-10 mm/yr). Comparing the profiles A and C supports the idea of slab sinking in the northwest Zagros and underthrusting in central Zagros (e.g., Mouthereau et al., 2012). Furthermore, our profiles reveal a distinguishing feature that sets apart northwest from southeast Zagros, a low-velocity corridor in the subcrustal depths beneath central and northern Zagros, observed in profiles B and C, respectively. Notably, this low-velocity corridor is absent in the southeast (profile A). These low velocities within the upper mantle typically indicate a weak mantle, potentially attributed to delaminated lithosphere beneath the NW Zagros (Manaman and Shomali 2010; Verdel et al., 2011; Shad Manaman et al., 2011; Mahmoodabadi et al., 2020), or it could be a result of mantle erosion due to partial melting (e.g., Al-lazki et al., 2003; Omrani et al., 2008), or it might be linked to near-solidus conditions in the uppermost mantle (e.g., Hearn and Ni, 1994). However, the relatively low convergence rate in the NW Zagros may have created a situation in which the thickened edge of the Zagros lithosphere delaminates from the lower crust. This lithospheric foundering or delamination has also been documented in different continental collision zones (e.g., Ren and Shen, 2008; Chen et al., 2017). Additionally, in profile D, which presents a cross-sectional view of the Zagros along its strike, a substantial low-Vs anomaly extending into the lower crustal depths beneath the southern Lurestan Arc, also known as the Izeh zone, separates the upper mantle of the central Zagros from the northernmost Zagros.

The observed thickening of the crust to the northeast of the MZRF (profiles A, B, and C) supports the previously proposed hypothesis of Arabian crust underthrusting beneath Central Iran, as was also outlined in previous studies (e.g., Paul et al., 2006, 2010; Mahmoodabadi et al., 2019; Irandoust et al., 2022). Furthermore, the progression of crustal deformation along the MZRF, where a transition from low-to-high Vs occurs, indicates the effective positioning of the suture zone between the SSZ and the MZRF. This suture zone is also evident in the horizontal sections (see Fig. 4, 40-50 km), separating the intense deformation of the Zagros from the aseismic and relatively flat rigid blocks found in the central Iranian Plateau (e.g., Engdahl et al., 2006; Talebian and Jackson, 2004). As we move towards central Iran, the crust thins (approximately 30-35 km), a characteristic also noted in other studies (e.g., Motaghi et al., 2012; Sodoudi et al., 2009; Wu et al., 2021). The positive anomalies observed beneath central Iran (profiles A, B and C) can be interpreted as remnants of the stable upper mantle that underlies the Central Iran microplate and was preserved during the closure of the Neo-Tethys Ocean.

Moving further to NE, profile B reveals a relatively thin crust (<40 km) beneath the north-central region of Iran. This thickness increases as we progress northward, reaching approximately 45 km across a broad area beneath the Kopet-Dagh Mountains. However, it is important to be cautious when interpreting the features in northeastern part of the study area due to the velocity smearing observed in synthetic test results (see Fig. 8). A similar variation in crustal thickness is evident to the east (profile A), indicating a thicker crust beneath the eastern part of the Lut block and the East basin. This is consistent with existence of the Sistan suture zone in the east (e.g., Tirrul., 1983). Notably, there are detectable high-velocity anomalies in the middle crust beneath CI (profiles A, B, and C), possibly associated with ophiolites at the surface or the remnants of the igneous intrusions. These shallow high-velocity anomalies have also been documented in previous studies by Motaghi et al. (2015) and Yaminifard et al. (2007).

Profiles C and E, from SW to NE and NNW to SSE, transect the Alborz, respectively. On both profiles the very low-Vs shallow layers of the SCB, representative of its sedimentary cover, are present in the coastal regions. Unlike the initial model and some previous studies (e.g., Hollingsworth et al., 2008; Tatar et al., 2007), no clear evidence of underthrusting SCB beneath the Alborz is observed. In a recent study by Motaghi et al. (2018), this absence of underthrusting is noted in the Alborz region. However, the underthrusting of the SCB beneath Alborz cannot be completely ruled out, as some studies (e.g., Rastgoo et al., 2018) attributed the low-velocity anomalies in western Alborz to post-collisional delamination of the lower part of the western Alborz lithosphere as a result of interaction between SCB and Alborz. In profile E, the crust beneath the Alborz is on average thinner than 40 km. The thin crust beneath the Alborz is suggested by some previous researches (Rodgers et al., 1997; Seber et al., 1997; Radjaee et al., 2010). However, in profile C, a thick crust beneath SW Alborz is evident, which is in agreement with results from Sodoudi et al. (2009). The low velocity anomaly in this area (profile C), which can be associated with a magma addition at the base of the crust, is coincident with the presence of volcanic material (Jackson et al., 2002) at the surface. Nevertheless, the presence of velocity smearing in this region, as observed in the results of synthetic tests (Fig. 8), implies a need for caution in interpreting this particular feature.

## 5. Conclusions

In this study, we have employed the adjoint noise tomography (ANT) technique to refine and create a more detailed 3D structural model for the Iranian Plateau, building upon an existing model (Kaviani et al., 2020) of shear-wave velocity in the crust and uppermost mantle derived from conventional ambient noise tomography. This enhanced model offers new insights into the Zagros collision zone and the crustal configuration beneath the Iranian plateau. The shear-wave velocity cross-sections spanning various parts of the Zagros collision zone, from the southeast to the northwest, reveal a distinct, tongue-shaped low-Vs structure, indicating the underthrusting of Arabian crust beneath central Iran. Notably, the angle of underthrusting in the northwest Zagros is steeper than in the southeast, supporting the idea of diachronous underthrusting development from northwest to southeast. Additionally, we have identified a subcrustal low-velocity corridor in the northwest and central Zagros, which we interpret as potential evidence for delamination of the thickened edge of the Zagros lithosphere from the lower crust. A notable local low-velocity feature beneath the southwest Alborz region may be linked to a magma addition at the base of the crust.

In summary, our new model refines and expands upon the features inherited from the initial model, unveiling new characteristics and contributing further to our understanding of the deep structural complexities beneath converging tectonic plates within the Iranian plateau.

**Acknowledgement**

The authors express their gratitude to the developers of the open-source software Specfem3D package by Komatitsch & Tromp (2002b, https://github.com/SPECFEM/specfem3d.git) and the version supporting AMD GPUs (https://github.com/ROCmSoftwarePlatform/SPECFEM3D). We extend our thanks to all data research centers (IRIS and NEIC) and networks, Iranian Seismological Center (IRSC), the Iranian National Seismic Network (ISNS), Dubai Seismological Network (DN), Global Seismograph Network - IRIS/IDA (II), Kandilli Observatory And Earthquake Research Institute (KO), Kuwait National Seismic Network (KW), Earthquake Monitoring Program of Oman (OM), Seismological Network of United Arab Emirates (UE) and Iraqi Seismic Observatory (IQ) that provided the data used in this study. Special appreciation is directed towards Prof. Dr. Qynia Liu from University of Toronto for her valuable comments, as well as Dr. Kai Wang from University of Science and Technology of China (USTC) and Dr. Ehsan Karkooti from International Institute of Earthquake Engineering and Seismology (IIEES) for their assistance in configuring the SPECFEM3D setup. AK acknowledges the generous support of a Goethe-University scholarship through the Department of Geosciences/Geography.

**Data Availability Statement**

*Data*: We used continuous waveforms from the Iranian Seismological Center (IRSC) and the Iranian National Seismic Network (ISNS), which are not publicly available. Data from the other stations are accessible via IRIS Data Services. The initial 3D shear-wave velocity model is available online through the IRIS (Incorporated Research Institutions for Seismology) Earth Model Collaboration (IRIS DMC 2011, http://ds.iris.edu/ds/products/emc-earthmodels/) with the label "Midd_East_Crust_1". The earthquake locations in Fig.1 were taken from the National Earthquake Information Center (NEIC) these are available on the U.S. Geological Survey (USGS) website at http://earthquake.usgs.gov.

*Software*: The data analysis was performed using the open-software SPECFEM3D code. The code used in this study (supporting the AMD GPUs) is available via the link: https://github.com/ROCmSoftwarePlatform/SPECFEM3D.

*Graphs*: The figures presented in the paper were generated using the PyGMT (https://www.pygmt.org) and Paraview (https://www.paraview.org).

Supplementary Material

# An improved high-resolution shear velocity model beneath the Iranian plateau using adjoint noise tomography

Abolfazl Komeazi[1*], Ayoub Kaviani[1], Farzam Yaminifard[2], Mohammad Tatar[2], Georg Rümpker[1,3]

*1. Institute of Geosciences, Goethe-University Frankfurt, Frankfurt, Germany*

*2. International Institute of Earthquake Engineering and Seismology (IIEES), Tehran, Iran*

*3. Frankfurt Institute for Advanced Studies, Frankfurt, Germany*

The Supplementary materials include seven figures (S1-S7) and one table (Table S1).

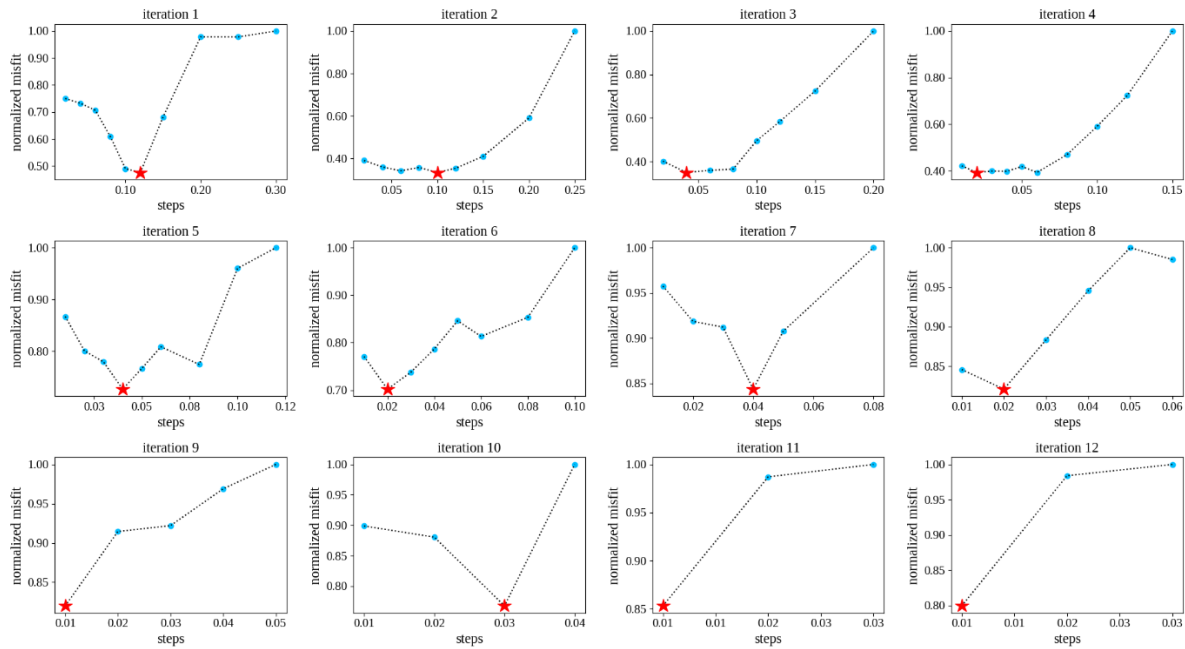

Figure S1. Results of the line search at each iteration, illustrating the change in misfit function values concerning the step length for the total misfit.

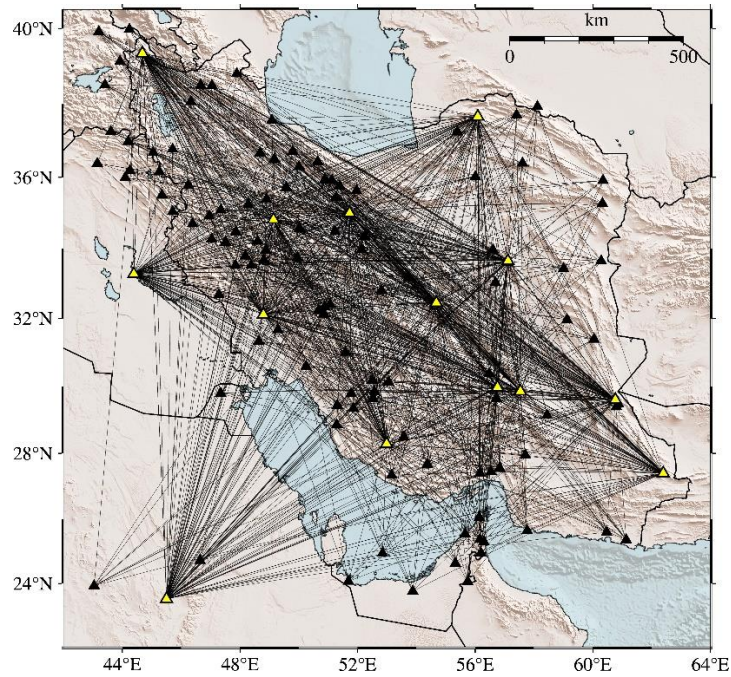

Figure S2. Location map of 14 source stations (marked with yellow triangles) and the corresponding source-receiver raypaths. These data are utilized for the selection of the optimal step length at each iteration.

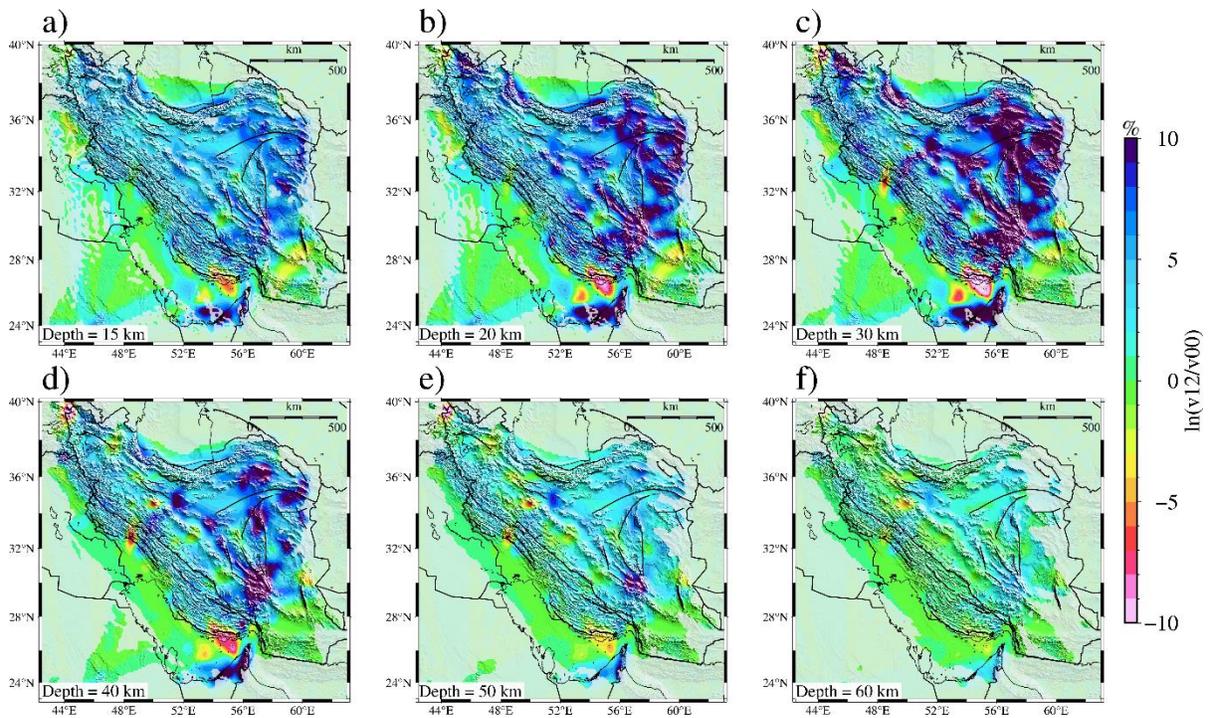

Figure S3. Difference in shear-wave velocity models (M00 and M12) shown in depth slices at 15, 20, 30, 40, 50 and 60 km. The relative perturbation of velocity to the mean velocity of each depth

slice (written on the lower left part of the map) are displayed to emphasize the changes in the shape and amplitude of anomalies. These maps are masked with respect to the misfit kernel values (see also Fig. S4 and S5).

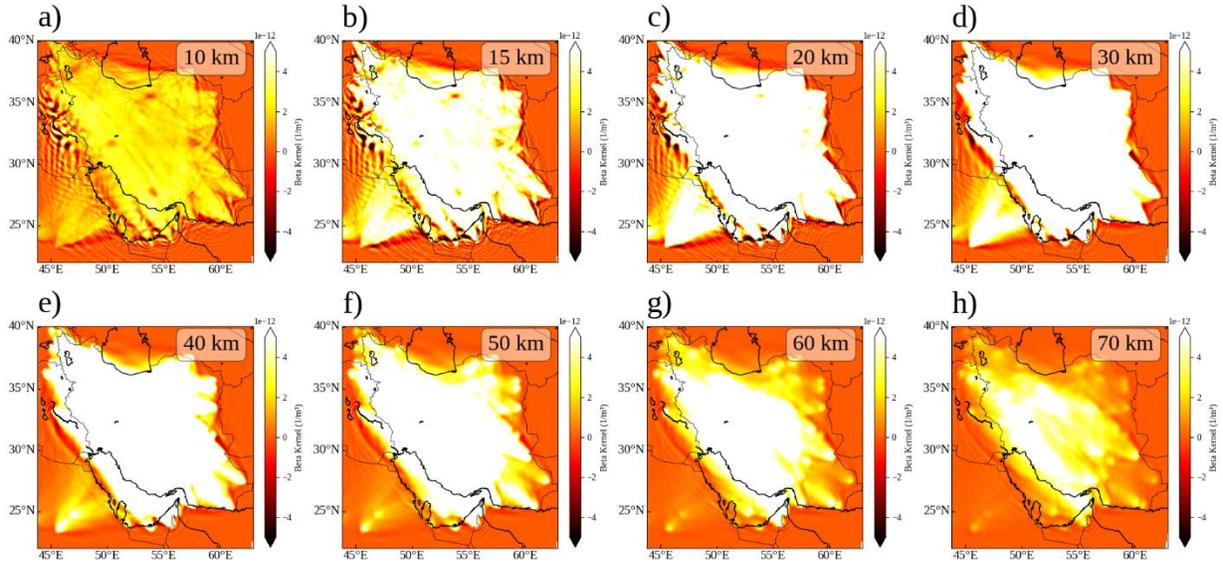

Figure S4. Volumetric sensitivity (misfit) kernels for shear wave velocity tomographic models shown at depths of 10, 15, 20, 30, 40, 50, 60 and 70 km.

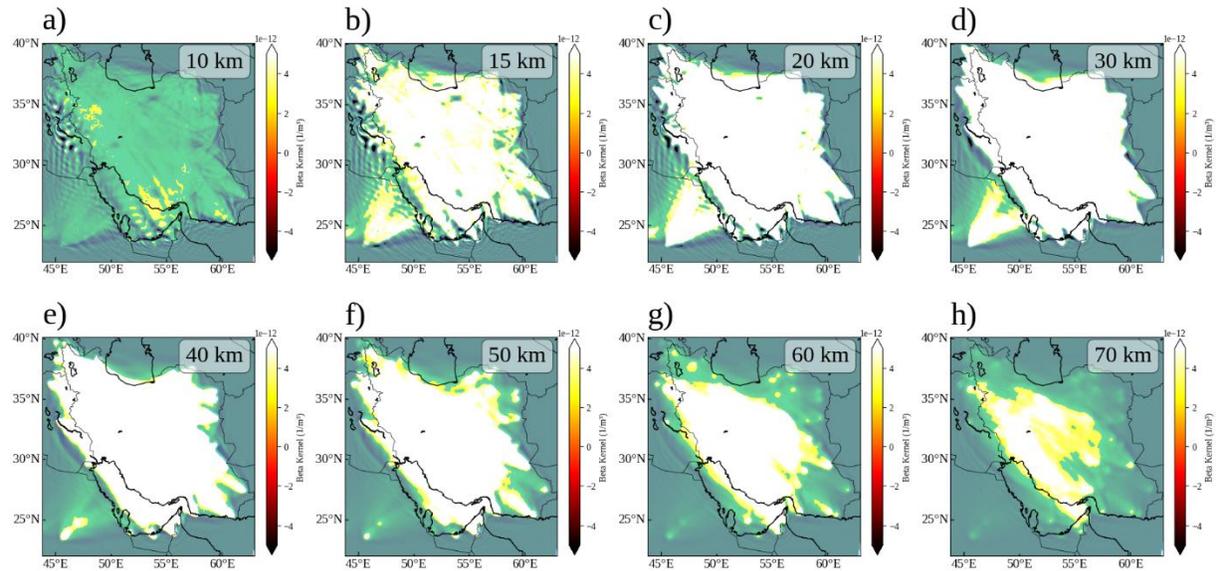

Figure S5. Same volumetric sensitivity kernels in Fig. S4 masked by a subjective threshold value of $K = 1.5 \times 10^{-13}\ m^{-3}$.

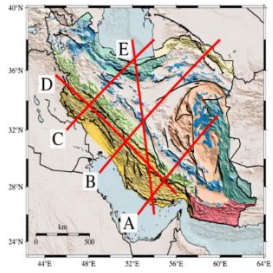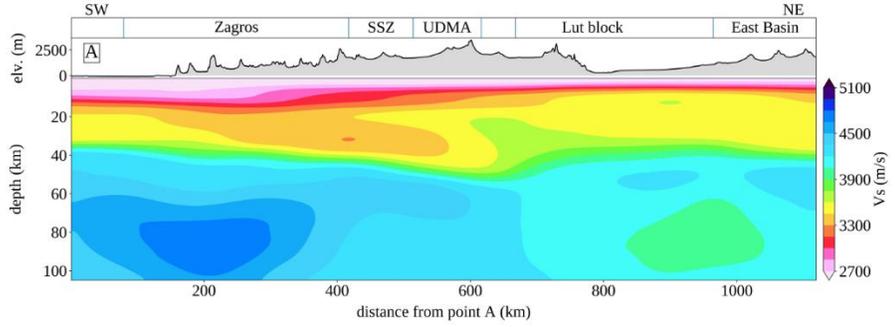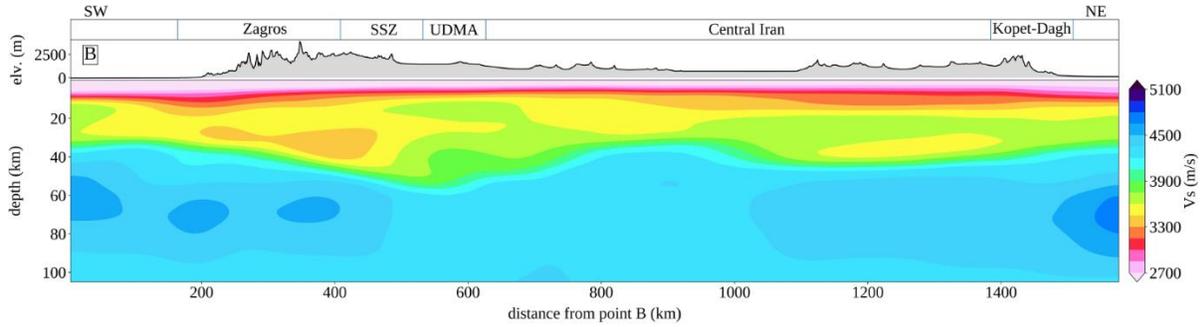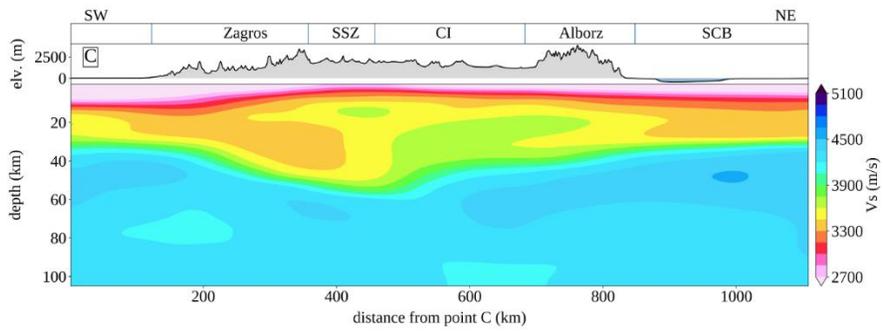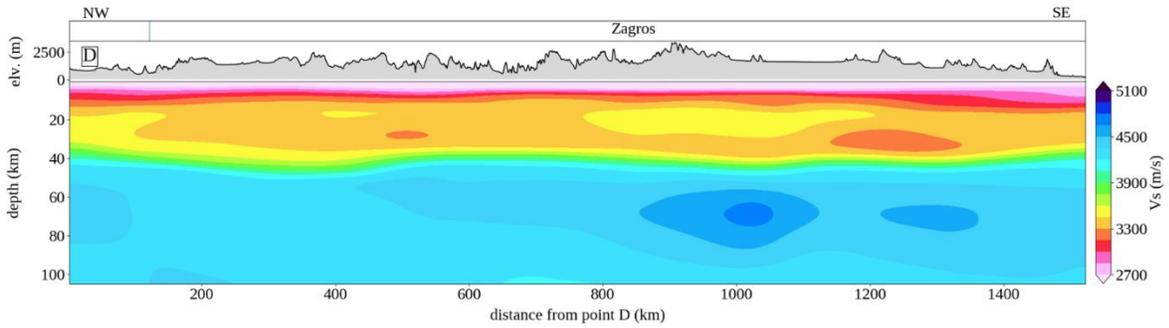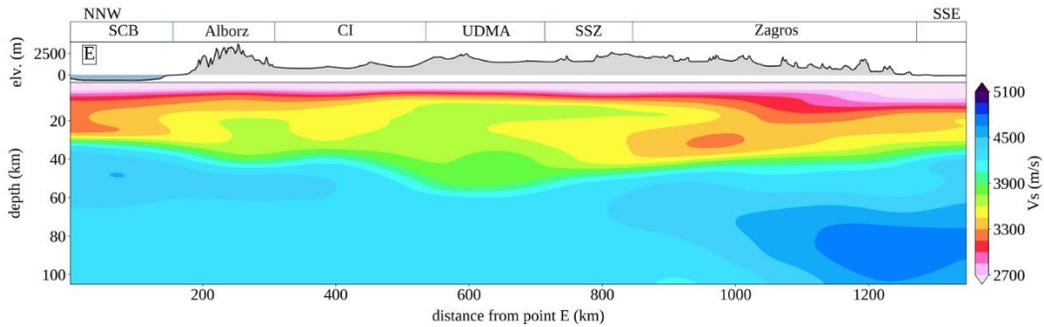

Figure S6. Vertical slices of the initial shear wave velocity model (M12) along profiles A-E across the Iranian Plateau. The positions of these profiles are indicated on the map in the top-left corner. Vertical exaggeration: 4.

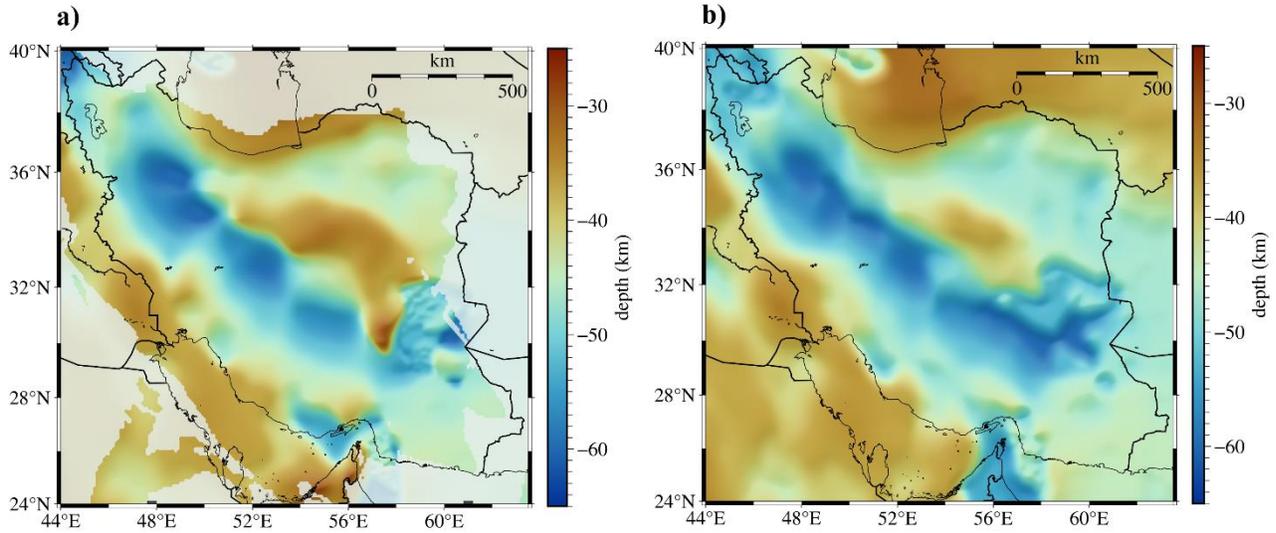

Figure S7. Moho depth maps for Vs = 4.2 iso-velocity surface – a) final model, b) initial model. Final model Moho map masked based on misfit kernel values (see also Fig. S4 and S5).

Table S1. Inversion parameter values across iterations: $\Delta t$ (time shift), $\Delta lnA$ (amplitude difference), and $CC_{min}$ (cross-correlation coefficient) are criteria for window selection. The two σ values (h and v) represent the horizontal and vertical radii of the Gaussian function used to smooth the misfit gradient, and α denotes the optimal step length selected at each iteration.

| Iteration Nummer | 1 | 2 | 3 | 4 | 5 | 6 | 7 | 8 | 9 | 10 | 11 | 12 |
|---|---|---|---|---|---|---|---|---|---|---|---|---|
| $\Delta t\ (s)$ | [−4.5, 4.5] | | | | | | | | | | | |
| $\Delta lnA$ | [−1.0, 1.0] | | | | | | | | | | | |
| $CC_{min}$ | 0.59 | 0.65 | | | 0.7 | | | 0.75 | | | 0.8 | |
| $\sigma(h,v)(km)$ | (15.0, 7.0) | | | | (10.0, 5.0) | | | | | | | |
| α | 0.12 | 0.1 | 0.04 | 0.02 | 0.04 | 0.02 | 0.04 | 0.02 | 0.01 | 0.03 | 0.01 | 0.01 |